\documentstyle[12pt]{article}
\def\lsim{\mathrel{\mathpalette\fun <}}
\def\gsim{\mathrel{\mathpalette\fun >}}
\def\fun#1#2{\lower3.6pt\vbox{\baselineskip0pt\lineskip.9pt
\ialign{$\mathsurround=0pt#1\hfil##\hfil$\crcr#2\crcr\sim\crcr}}}

\newcommand{\be}{\begin{eqnarray}}
\newcommand{\ee}{\end{eqnarray}}

\input{epsfig.sty}

\begin{document}

\Huge{\noindent{Istituto\\Nazionale\\Fisica\\Nucleare}}

\vspace{-3.9cm}

\Large{\rightline{Sezione SANIT\`{A}}}
\normalsize{}
\rightline{Istituto Superiore di Sanit\`{a}}
\rightline{Viale Regina Elena 299}
\rightline{I-00161 Roma, Italy}

\vspace{0.65cm}

\rightline{INFN-ISS 97/13}
\rightline{November 1997}

\vspace{2cm}

\begin{center}

\large{\bf ISGUR-WISE FORM FACTORS OF HEAVY BARYONS WITHIN A
LIGHT-FRONT CONSTITUENT QUARK MODEL}

\vspace{1cm}

\large{F. Cardarelli and S. Simula}

\vspace{0.5cm}

\normalsize{Istituto Nazionale di Fisica Nucleare, Sezione Sanit\`{a},
\\ Viale Regina Elena 299, I-00161 Roma, Italy}

\end{center}

\vspace{1cm}

\begin{abstract}

\indent The space-like elastic form factors of baryons containing a
heavy quark are investigated within a light-front constituent quark
model in the limit of infinite heavy-quark mass, adopting a
gaussian-like ans\"atz for the three-quark wave function. The results
obtained for the Isgur-Wise form factors corresponding both to a
spin-0 and a spin-1 light spectator pair are presented. It is found
that the Isgur-Wise functions depend strongly on the baryon structure,
being sharply different in case of diquark-like or collinear-type
configurations in the three-quark system. It is also shown that the
relativistic effects lead to a saturation property of the form factors
as a function of the baryon size. Our results are compared with those
of different models as well as with recent predictions from $QCD$ sum
rules and lattice $QCD$ simulations; the latter ones seem to suggest
the dominance of collinear-type configurations, in which the
heavy-quark is sitting close to the center-of-mass of the light quark
pair. 

\end{abstract}

\vspace{1cm}

\noindent PACS numbers: 12.38.Lg; 12.39.Ki; 13.40.Gp; 14.20.Mr

\vspace{1cm}

\noindent Keywords: heavy baryons; relativistic quark model;
electroweak form factors.

\newpage

\pagestyle{plain}

\indent The investigation of heavy-hadron decays can provide relevant
information on fundamental parameters of the Standard Model, like the
Cabibbo-Kobayashi-Maskawa mixing angles and the quark masses. As it is
well known, in case of hadrons containing a single heavy quark ($Q$),
the complexity of the theoretical analysis is largely simplified by
the Heavy Quark Symmetry ($HQS$) \cite{IW,HQET,NEU94}, which is a
spin-flavour symmetry shared by (but not manifest in) the $QCD$
Lagrangian when the limit of infinite heavy-quark mass ($m_Q \to
\infty$) is considered. The $HQS$, which is only softly broken by
terms of order $\Lambda_{QCD} / m_Q$, allows to derive several
model-independent relations among hadronic properties; in particular,
it requires that, when $m_Q \to \infty$, all the non-perturbative
strong physics relevant for exclusive electroweak processes is
contained in few universal form factors, known as the Isgur-Wise
($IW$) functions. The latter however cannot be predicted by the $HQS$,
for the full knowledge of the non-perturbative structure of heavy
hadrons is required. One has therefore to resort to models in order to
make quantitative predictions, provided the $HQS$ relations are
checked in any particular model for internal consistency.

\indent The aim of this letter is to investigate the $IW$ form factors
of heavy baryons using a relativistic constituent ($CQ$) quark model
formulated on the light-front ($LF$). Our $CQ$ model (see Refs.
\cite{CAR,SIM96}), which properly incorporates boost effects and the
relativistic composition of $CQ$ spins, has been already applied to
the evaluation of the $IW$ form factor in case of the ground-state of
heavy-light mesons. It has been shown \cite{SIM96} that in a wide
range of values of the recoil ($\omega - 1$), where $\omega \equiv v
\cdot v'$ is the product of the initial ($v$) and final ($v'$) hadron
four-velocities, the calculated $IW$ function $\xi^{(IW)}(\omega)$
exhibits a moderate dependence upon the choice of the heavy-meson wave
function; moreover, the slope of the $IW$ form factor at the
zero-recoil point, $\rho^2 \equiv - [d\xi^{(IW)}(\omega) /
d\omega]_{\omega = 1}$, has been found to be remarkably increased by
relativistic effects and quantitatively close to the value $\rho^2
\simeq 1$ \cite{SIM96} (cf. also Ref. \cite{CW94}). In this letter,
the baryon $IW$ form factors are obtained from the electromagnetic
(e.m.) elastic form factors of heavy baryons, corresponding both to a
spin-0 and a spin-1 light spectator pair, calculated for space-like
values of the squared four-momentum transfer $q^2 \equiv q \cdot q <
0$. Though we limit ourselves to a gaussian-like ans\"atz for the
baryon wave function, an interesting feature of the $IW$ form factors
emerges. Indeed, it will be shown that the baryon $IW$ functions are
sharply different in case of diquark-like or collinear-type
configurations in the three-quark system. In particular, the slope of
the $IW$ form factors at the zero-recoil point is remarkably smaller
for collinear-type configurations, in which the heavy quark is sitting
close to the center-of-mass of the two light spectator quarks. It is
also found that the relativistic effects lead to a saturation property
of the $IW$ form factors as a function of the baryon size.

\indent To begin with, let us consider a $Q (q q')$ baryon composed by
a heavy quark $Q$ with constituent mass $m_Q$ and a partner light-quark
pair $(q q')$ with constituent masses $m_q$ and $m_{q'}$. Within the
$LF$ formalism (cf., e.g., Refs. \cite{CAR,SIM96}) the hadron wave
functions are eigenvectors of a mass operator, e.g. ${\cal{M}} = M_0 +
{\cal{V}}$, and of the non-interacting angular momentum operators
$j^2$ and $j_n$, where the vector $\hat{ n} = (0,0,1)$ defines the
spin quantization axis. The operator $M_0$ is the free mass and the
interaction term ${\cal{V}}$ is a Poincar\'e invariant. For baryons
one has $M_0^2 = \sum_{i = Q, q, q'} (k_{i \perp}^2 + m_i^2) / \xi_i$,
where $\xi_i = p_i^+ / P^+$ and $\vec{k}_{i \perp} = \vec{p}_{i \perp}
- \xi_i \vec{P}_{\perp}$ are the intrinsic $LF$ variables. The
subscript $\perp$ indicates the projection perpendicular to the spin
quantization axis and the {\em plus} component of a four-vector $p
\equiv (p^0, \vec{p})$ is given by $p^+ = p^0 + \hat{n} \cdot \vec{p}$;
finally, $\tilde{P} \equiv (P^+, \vec{P}_{\perp}) = \tilde{p}_Q +
\tilde{p}_q + \tilde{p}_{q'}$ is the $LF$ baryon momentum and
$\tilde{p}_i$ the quark one. In terms of the longitudinal momentum
$k_{in}$, related to the variable $\xi_i$ by $k_{in} = {1 \over 2} ~
\left [ \xi_i M_0 - (k_{i \perp}^2 + m_i^2) / \xi_i M_0 \right ]$, the
free mass  operator acquires a familiar form, viz.
 \be
    M_0 = \sum_{i = Q, q, q'} ~ \sqrt{m_i^2 + |\vec{k}_i|^2 } \equiv
    \sum_{i = Q, q, q'} ~ E_i
    \label{eq:free_mass} 
 \ee
with $\vec{k}_i \equiv ( \vec{k}_{i \perp}, k_{in})$. Disregarding
for simplicity the colour and flavour degrees of freedom and limiting
ourselves to $S$-wave baryons, the $LF$ wave function can be written as
 \be
    \langle \{ \xi_i \vec{k}_{i \perp}; \nu_i \}| \Psi_{S_{q q'}}^{J
    \mu} \rangle = \sqrt{{E_Q E_q E_{q'} \over M_0 \xi_Q \xi_q
    \xi_{q'}}} \sum_{\{\nu'_i \}} \langle \{ \nu_i \} |
    {\cal{R}}^{\dag}(\{ \vec{k}_i; m_i \}) | \{ \nu'_i \} \rangle ~
    \langle \{ \vec{k}_i; \nu'_i \}| \chi_{S_{q q'}}^{J \mu} \rangle
    \label{eq:LF_wf} 
 \ee 
where $J$ and $\mu$ are the total angular momentum and its projection;
$S_{q q'}$ is the spin of the light quark pair ($q q'$); the curly
braces $\{ ~~ \}$ mean a list of indices corresponding to $i = Q, q,
q'$; $\nu'_i$ is the third component of the $CQ$ spin; ${\cal{R}}(\{
\vec{k}_i; m_i \}) \equiv \prod_{j = Q, q, q'} R_M(\vec{k}_j, m_j)$
with $R_M(\vec{k}_j, m_j)$ being the (generalized) Melosh rotation.
The canonical heavy-baryon wave function $\langle \{ \vec{k_{i}};
\nu'_i \}| \chi_{S_{q q'}}^{J \mu} \rangle$ is given by
 \be
    \langle \{ \vec{k}_i; \nu'_i \}| \chi_{S_{q q'}}^{J \mu} \rangle =
    w_{(Q q q')}(\vec{p}, \vec{k}) \cdot \Phi_{S_{q q'}}^{J \mu} ( \{
    \nu'_i \}) 
    \label{eq:ET_wf}
 \ee
where $\vec{p} = [(m_q + m_{q'}) \vec{k}_Q - m_Q (\vec{k}_q +
\vec{k}_{q'})] / (m_Q + m_q + m_{q'}) =$ $\vec{k}_Q = - \vec{k}_q -
\vec{k}_{q'}$ and $\vec{k} = (m_{q'} \vec{k}_q - m_q \vec{k}_{q'}) /
(m_q + m_{q'})$ are the Jacobi coordinates for the $Q (q q')$ system;
$w_{(Q q q')}$ is the $S$-wave radial wave function and the spin
function $\Phi_{S_{q q'}}^{J \mu}(\{ \nu'_i \})$ is defined as
 \be
    \Phi_{S_{q q'}}^{J \mu}(\{ \nu'_i \}) = \sum_{\sigma} \langle
    {1 \over 2} \nu'_Q S_{q q'} \sigma | J \mu \rangle ~ \langle {1
    \over 2} \nu'_q {1 \over 2} \nu'_{q'}|S_{q q'} \sigma \rangle 
    \label{eq:spin_wf} 
 \ee  
For a fixed value of $S_{q q'}$ ($= 0, 1$) the $LF$ wave functions
(\ref{eq:LF_wf}) with $J = S_{q q'} \pm 1/2$ belong to multiplets of
the $HQS$ (see Ref. \cite{IW}(c)), namely to singlets for $S_{q q'} =
0$ ($J = 1/2$) and doublets for $S_{q q'} = 1$ ($J = 1/2, 3/2$).
Finally, the wave function (\ref{eq:ET_wf}) is normalized as: $\sum_{\{
\nu_i \}} \int \{ d\vec{k}_i \} ~ \delta(\vec{k}_Q + \vec{k}_q +
\vec{k}_{q'}) ~ \left |\langle \{ \vec{k}_i, \nu_i \} | \chi_{S_{q
q'}}^{J \mu} \rangle \right|^2 = 1$, which implies $\int d\vec{p}
d\vec{k} ~ |w_{(Q q q')}(\vec{p}, \vec{k})|^2 = 1$. In what follows we
will limit ourselves to a gaussian ans\"atz for $w_{(Q q q')}(\vec{p},
\vec{k})$, viz.
 \be
    w_{(Q q q')}(\vec{p}, \vec{k}) = ({1 \over \pi \alpha_p})^{3/2}
    e^{-|\vec{p}|^2 / 2 \alpha_p^2} ~ ({1 \over \pi \alpha_k})^{3/2}
    e^{-|\vec{k}|^2 / 2 \alpha_k^2}
    \label{eq:gaussian}
 \ee
where $\alpha_p$ and $\alpha_k$ are the harmonic oscillator parameters
governing the average value of the internal momenta, namely $\langle
|\vec{p}|^2 \rangle = {3 \over 2} \alpha_p^2$ and $\langle |\vec{k}|^2
\rangle = {3 \over 2} \alpha_k^2$.

\indent After the description of the structure of our $LF$ three-quark
wave function, we now briefly remind (see Ref. \cite{IW}(c)) that
in the heavy-quark limit all the relevant electroweak baryon form
factors reduce to one $IW$ function, $\xi^{(0)}(\omega)$, in case of
the singlets with $S_{q q'} = 0$ and to two $IW$ form factors,
$\xi^{(1)}(\omega)$ and $\zeta^{(1)}(\omega)$, in case of the doublets
with $S_{q q'} = 1$. In this letter the $IW$ form factors are
calculated through the matrix elements of the e.m. vector current
$\bar{Q} \gamma^{\rho} Q$ between $J = 1/2$ baryons, viz.
 \be
    {\cal{V}}_{\mu' \mu}^{\rho}(S_{q q'}) & \equiv & \langle
    \Psi_{S_{q q'}}^{1/2 \mu'} | \bar{Q} \gamma^{\rho} Q | \Psi_{S_{q
    q'}}^{1/2 \mu} \rangle = F_1^{(S_{q q'})}(\omega) ~ \bar{u}(P',
    \mu') \gamma^{\rho} u(P, \mu) + \nonumber \\
    & + & \left [ F_2^{(S_{q q'})}(\omega) ~ v^{\rho} + F_3^{(S_{q
    q'})}(\omega) ~ {v'}^{\rho} \right ] ~ \bar{u}(P', \mu') u(P, \mu)
    \label{eq:vector}
 \ee
where $u(P, \mu)$ is a Dirac spinor (normalized as $\bar{u} u = 1$), $P
= M v$ and $P' = P + q = M v'$, with $M$ being the heavy-baryon
mass. Since we are considering an elastic process, the four-momentum
transfer squared is given by $q^2 = 2 M^2 \cdot (1 - \omega)$ and
has space-like values ($q^2 \leq 0$). A formula similar to Eq.
(\ref{eq:vector}) can be written for the matrix elements of the
axial-vector current $\bar{Q} \gamma^{\rho} \gamma_5 Q$, namely
 \be
    {\cal{A}}_{\mu' \mu}^{\rho}(S_{q q'}) & \equiv & \langle
    \Psi_{S_{q q'}}^{1/2 \mu'} | \bar{Q} \gamma^{\rho} \gamma_5 Q |
    \Psi_{S_{q q'}}^{1/2 \mu} \rangle = G_1^{(S_{q q'})}(\omega) ~
    \bar{u}(P', \mu') \gamma^{\rho} \gamma_5 u(P, \mu) + \nonumber \\
    & + & \left [ G_2^{(S_{q q'})}(\omega) ~ v^{\rho} + G_3^{(S_{q
    q'})}(\omega) ~ {v'}^{\rho} \right ] ~ \bar{u}(P', \mu') \gamma_5
    u(P, \mu)
    \label{eq:axial}
 \ee
In the heavy-quark limit one has \cite{IW}(c)
 \be
    \mbox{lim}_{m_Q \to \infty} F_1^{(0)}(\omega) & = &
    \mbox{lim}_{m_Q \to \infty} G_1^{(0)}(\omega) = \xi^{(0)}(\omega)
    \label{eq:F1G1_0} \\
    \mbox{lim}_{m_Q \to \infty} F_2^{(0)}(\omega) & = &
    \mbox{lim}_{m_Q \to \infty} F_3^{(0)}(\omega) = 0
    \label{eq:F2F3_0} \\
    \mbox{lim}_{m_Q \to \infty} G_2^{(0)}(\omega) & = &
    \mbox{lim}_{m_Q \to \infty} G_3^{(0)}(\omega) = 0
    \label{eq:G2G3_0} \\
    \mbox{lim}_{m_Q \to \infty} F_1^{(1)}(\omega) & = &
    \mbox{lim}_{m_Q \to \infty} G_1^{(1)}(\omega) = - {1 \over 6}
    \left [ 2 \omega \xi^{(1)}(\omega) + (\omega - 1)
    \zeta^{(1)}(\omega) \right ]
    \label{eq:F1G1_1} \\
    \mbox{lim}_{m_Q \to \infty} F_2^{(1)}(\omega) & = &
    \mbox{lim}_{m_Q \to \infty} F_3^{(1)}(\omega) = {2 \over 3} ~
    \xi^{(1)}(\omega)
    \label{eq:F2F3_1} \\
    \mbox{lim}_{m_Q \to \infty} G_2^{(1)}(\omega) & = & -
    \mbox{lim}_{m_Q \to \infty} G_3^{(1)}(\omega) = {2 \over 3} ~
    \left [ \xi^{(1)}(\omega) + \zeta^{(1)}(\omega) \right ]
    \label{eq:G2G3_1}
 \ee 
where the $IW$ form factors $\xi^{(0)}(\omega)$ and $\xi^{(1)}(\omega)$
must satisfy the model-independent normalization $\xi^{(0)}(1) =
\xi^{(1)}(1) = 1$ at the zero-recoil point $\omega = 1$, whereas the
normalization $\zeta^{(1)}(1)$ is not known {\em a priori}. In what
follows we will choose a reference frame where $q^+ \equiv q^0 +
\hat{n} \cdot \vec{q} = 0$, which allows to suppress the contribution
arising from the so-called $Z$-graph (pair creation from the vacuum)
at any value of the heavy-quark mass.

\indent All the hadronic form factors corresponding to a conserved
current can always be expressed in terms of the matrix elements of the
{\em plus} component of the current. As a matter of fact, adopting a
reference frame where $q^+ = 0$, from Eq. (\ref{eq:vector}) one easily
gets
 \be
    F_1^{(S_{q q'})}(\omega) & = & {1 \over 2}
    \mbox{Tr}\left\{ {{\cal{V}}^+(S_{q q'}) \over 2P^+} \right\} +
    {M \over \sqrt{-q^2}} \mbox{Tr}\left\{ {{\cal{V}}^+(S_{q q'})
    i \sigma_2 \over 2P^+} \right\}
    \label{eq:F1} \\
    F_2^{(S_{q q'})}(\omega) & = & F_3^{(S_{q q'})}(\omega) = - 
    {M \over 2 \sqrt{-q^2}} \mbox{Tr}\left\{ {{\cal{V}}^+(S_{q q'})
    i \sigma_2 \over 2P^+} \right\}
    \label{eq:F2F3}
 \ee
On the contrary, from Eq. (\ref{eq:axial}) one has
 \be
    G_1^{(S_{q q'})}(\omega) & = & {1 \over 2} \mbox{Tr}\left\{
    {{\cal{A}}^+(S_{q q'}) \sigma_3 \over 2P^+} \right\}
    \label{eq:G1} \\
    G_2^{(S_{q q'})}(\omega) + G_3^{(S_{q q'})}(\omega) & = & -
    {M \over \sqrt{-q^2}} \mbox{Tr}\left\{ {{\cal{A}}^+(S_{q q'})
    \sigma_1 \over 2P^+} \right\}
    \label{eq:G2plusG3} \\
    G_3^{(S_{q q'})}(\omega) - G_2^{(S_{q q'})}(\omega) & = & {2M^2
    \over -q^2} \left[ \mbox{Tr}\left\{ {{\cal{A}}^+(S_{q q'})
    \sigma_3 \over 2P^+} \right\} - \right. \nonumber \\ 
    & & \left. {P^+ \over M} \mbox{Tr}\left\{ {{\cal{A}}_{\perp}(S_{q
    q'}) \sigma_1 \over 2P^+} \right\} \right]
    \label{eq:G3minusG2}
 \ee
with $P^+ = \sqrt{M^2 - q^2 / 4}$. Therefore, the evaluation of the
form factors $G_2^{(S_{q q'})}(\omega)$ and $G_3^{(S_{q q'})}(\omega)$
requires also the use of the matrix element of the transverse
component of the (non-conserved) axial current.

\indent We have evaluated the right-hand side of Eqs.
(\ref{eq:F1}-\ref{eq:G3minusG2}) using the three-quark wave function
given by Eqs. (\ref{eq:LF_wf}-\ref{eq:spin_wf}) adopting the gaussian
ans\"atz (\ref{eq:gaussian}). The numerical integrations, involving
six-dimensional integrals, have been performed through a
well-established Monte Carlo procedure \cite{VEGAS}. The heavy-quark
limit ($m_Q \to \infty$) is obtained by increasing the value of the
heavy-quark mass until full convergence of the calculated form factors
is reached. We have found that all the $HQS$ relations
(\ref{eq:F1G1_0}-\ref{eq:G2G3_1}) are fulfilled at any value of
$\omega$, except Eq. (\ref{eq:G2G3_1}) at the zero-recoil point $\omega
= 1$, because of the occurrence of a divergence at $q^2 = 0$ in the
right-hand side of Eq. (\ref{eq:G3minusG2}). The understanding and
elimination of such a divergence is outside the scope of the present
letter and will be addressed in a future work \cite{CS}. Here, we will
limit ourselves to note that within the $LF$ formalism the transverse
components of the current can contain the so-called instantaneous
propagation terms, which are instead absent in the {\em plus}
component. Thus, the $IW$ form factors $\xi^{(0)}(\omega)$,
$\xi^{(1)}(\omega)$ and $\zeta^{(1)}(\omega)$ will be determined only
via the matrix elements of the {\em plus} component of the (conserved)
vector current, viz.
 \be
    \xi^{(0)}(\omega) & = & {1 \over 2} \mbox{lim}_{m_Q \to \infty}
    \mbox{Tr}\left\{ {{\cal{V}}^+(0) \over 2P^+} \right\}
    \label{eq:csi0} \\
    \xi^{(1)}(\omega) & = & - {3 \over 4 \omega \sqrt{2(\omega - 1)}}
    \mbox{lim}_{m_Q \to \infty} \mbox{Tr}\left\{ {{\cal{V}}^+(1)
    i \sigma_2 \over 2P^+} \right\} 
    \label{eq:csi1} \\
    \zeta^{(1)}(\omega) & = & - {3 \over \omega - 1} \mbox{lim}_{m_Q
    \to \infty}\left [ \mbox{Tr}\left\{ {{\cal{V}}^+(1) \over 2P^+}
    \right\} + {3 \over 2 \sqrt{2(\omega - 1)}} \mbox{Tr}\left\{
    {{\cal{V}}^+(1) i \sigma_2 \over 2P^+} \right\} \right ]
    \label{eq:zeta1}
 \ee
The explicit expressions of the traces appearing in Eqs.
(\ref{eq:csi0}-\ref{eq:zeta1}) will be reported elsewhere \cite{CS}.
Here, we point out that the absolute normalizations $\xi^{(0)}(1) =
\xi^{(1)}(1) = 1$ are recovered in the right-hand side of Eqs.
(\ref{eq:csi0}-\ref{eq:csi1}) thanks to the normalization of the
canonical wave function (\ref{eq:ET_wf}) and to the $SU(2)$
Clebsh-Gordan algebra; moreover, for the same reasons the right-hand
side of Eq. (\ref{eq:zeta1}) behaves regularly when $\omega \to 1$.

\indent In our calculations, besides the masses $m_q$ and $m_{q'}$ of
the spectator quarks, there are two more parameters, $\alpha_p$ and
$\alpha_k$, appearing in the radial wave function (\ref{eq:gaussian}).
Instead of them we now introduce two other parameters, which are
combinations of $\alpha_p$ and $\alpha_k$, inspired by the fact that
in the heavy-quark limit ($m_Q \to \infty$) the canonical wave
function (\ref{eq:ET_wf}) with the gaussian ans\"atz
(\ref{eq:gaussian}) implies a root mean square radius of the baryon
equal to $\sqrt{<r^2>_B} \equiv \sqrt{3 \over 2} ~ \alpha^{-1}$, where
 \be
    \label{eq:alpha}
    \alpha & \equiv & {\alpha_p \over \sqrt{2 + {1 + \eta^2 \over (1 +
    \eta)^2} \beta^2}} \\
    \label{eq:beta}
    \beta & \equiv & {\alpha_p \over \alpha_k} = {<|\vec{p}|^2> \over
    <|\vec{k}|^2>}
 \ee
with $\eta \equiv m_q / m_{q'}$ being the light $CQ$ mass ratio. Thus,
the parameter $\alpha$ (Eq. (\ref{eq:alpha})) governs the (canonical)
baryon size, whereas $\beta$ (Eq. (\ref{eq:beta})) is the ratio of the
average internal momenta. Let us first focus on $\alpha$ and consider
only the form factor $\xi^{(0)}(\omega)$ for sake of simplicity. In
the non-relativistic limit the {\em charge} radius of
$\xi^{(0)}(\omega)$, $\rho_{(0)}^2 \equiv - [d \xi^{(0)}(\omega) /
d\omega]_{\omega = 1}$, is proportional to the square of the
(canonical) baryon size $<r^2>_B$, so that when $\alpha \to \infty$ one
should have $\rho_{(0)}^2 \to 0$. However, a well-known feature of the
relativistic effects is the delocalization of the position of the
light particles, triggered when the baryon size becomes smaller than
$\sqrt{\lambda_q^2 + \lambda_{q'}^2}$, where $\lambda_{q(q')} \equiv
1/ m_{q(q')}$ is the $CQ$ Compton wavelength. Therefore, we expect the
slope $\rho_{(0)}^2$ to be a monotonically decreasing function of
$\alpha$, and moreover, when $\alpha \gsim \alpha_{sat}$ with
 \be
   \alpha_{sat} \equiv \sqrt{3 \over 2} ~{m_q m_{q'} \over \sqrt{m_q^2
   + m_{q'}^2}} ~ ,
   \label{eq:alpha_sat}
 \ee
the slope $\rho_{(0)}^2$ should saturate and become independent of the
baryon size, being dominated by relativistic effects\footnote{However,
the saturation value of the slope $\rho_{(0)}^2$ still depends on the
particular functional form adopted for the radial wave function
$w_{(Q q q')}(\vec{p}, \vec{k})$.}. These expectations, including the
dependence of $\alpha_{sat}$ (Eq. (\ref{eq:alpha_sat})) upon the
spectator $CQ$ masses, are {\em fully} confirmed by explicit
calculations of the slope $\rho_{(0)}^2$ in the $(u, d, s, c)$
spectator sector. In Fig. 1 we have reported the results obtained for
the form factor $\xi^{(0)}(\omega)$ for various values of $\alpha$ at
fixed $\beta$. It can clearly be seen that in a wide range of values
of $\omega$, including the range of interest for the $b \to c$
transition, the form factor itself {\em saturates} for $\alpha \gsim
\alpha_{sat}$. Similar results hold as well at any value of $\beta$
and for the spin-1 form factors $\xi^{(1)}(\omega)$ and
$\zeta^{(1)}(\omega)$, so that for $\alpha \gsim \alpha_{sat}$ the
baryon $IW$ form factors becomes almost independent of the parameter
$\alpha$, i.e. of the baryon size. Since the (flavour-independent)
confinement scale yields $\alpha \sim 0.4 ~ GeV$ (cf. Ref.
\cite{SIM96}), from Eq. (\ref{eq:alpha_sat}) it follows that in case
of spectator $CQ$ masses in the $(u, d, s)$ sector the baryon $IW$
form factors are expected to be almost independent of the baryon size
and therefore to be in our gaussian model function of $\beta$ and
$\eta$ only.

\indent As far as the parameter $\beta$ is concerned, we have
sketched in Fig. 2 two types of three-quark spatial configurations
corresponding to the limiting cases $\beta << 1$ (Fig. 2(a)) and
$\beta >> 1$ (Fig. 2(b)). In the heavy-quark limit ($m_Q \to \infty$)
the internal momentum $\vec{p}$ describes the motion of the
center-of-mass of the spectator pair with respect to the heavy-quark
sitting on the center-of-mass of the baryon, i.e. the Jacobian
momentum $\vec{p}$ [$\vec{k}$] is conjugate to the spatial variable
$\vec{R}_{Q(q q')} \equiv$ $ \vec{r}_Q - (m_q \vec{r}_q + m_{q'}
\vec{r}_{q'}) / (m_q + m_{q'})$ [$\vec{r}_q - \vec{r}_{q'}$], where
$\vec{r}_j$ is the position of the quark $j$. Since $\beta = \langle
|\vec{r}_q - \vec{r}_{q'}|^2 \rangle / \langle |\vec{R}_{Q(q q')}|^2
\rangle$, the range of values $\beta \lsim 1$ is characterized by
values of $|\vec{R}_{Q(q q')}|$ larger than the average relative
distance $|\vec{r}_q - \vec{r}_{q'}|$ between the light spectator
quarks; this kind of configuration resembles a diquark-like cluster
(see Fig. 2(a) in the limiting case $\beta << 1$). On the contrary,
when $\beta > 1$, one has on average $|\vec{R}_{Q(q q')}| < |\vec{r}_q
- \vec{r}_{q'}|$, resembling now a collinear-type configuration, where
the heavy-quark is sitting close to the center-of-mass of the
spectator $CQ$ pair (see Fig. 2(b) in the limiting case $\beta >> 1$).
In Fig. 3 the $IW$ form factors $\xi^{(0)}(\omega)$,
$\xi^{(1)}(\omega)$ and $\zeta^{(1)}(\omega)$ are shown for various
values of the parameter $\beta$ in the saturation region $\alpha >
\alpha_{sat}$. It can be seen that the $IW$ form factors depend
strongly on the baryon structure, being sharply different in case of
diquark-like ($\beta \lsim 1$) or collinear-type ($\beta > 1$)
configurations in the three-quark system. This result is at variance
with what happens in the heavy-light meson case, where the sensitivity
of the $IW$ form factor to the choice of the meson wave function is
moderate \cite{SIM96}. Therefore, in case of heavy-light baryons the
investigation of the $IW$ functions appears to be very appealing
in order to shed light on the non-perturbative aspects of $QCD$. The
dynamical determination of the parameter $\beta$, which should depend
upon the spin of the spectator quark pair, is clearly called for and,
in this respect, calculations based on baryon wave functions derived
from spectroscopic $CQ$ models are in progress \cite{CS}\footnote{In
Ref. \cite{CS} we will address in details also the dependence of the
$IW$ functions on the spectator $CQ$ mass ratio $\eta = m_q /
m_{q'}$.}. In Fig. 3(a) our predictions for $\xi^{(0)}(\omega)$ are
compared with existing results from the diquark model of Ref.
\cite{GK93}, the $MIT$ bag model of Ref. \cite{SZ93}, the Skyrme model
of Ref. \cite{JMW96}, the $QCD$ sum rule technique of Ref.
\cite{YAK96} and with recent lattice $QCD$ simulations \cite{UKQCD}.
It can be seen that, in qualitative agreement with our results, the
diquark model predicts the steepest $IW$ function $\xi^{(0)}(\omega)$,
while both the lattice points and the $QCD$ sum rule results indicate
a much softer form factor, suggesting the dominance of collinear-type
configurations in the structure of heavy-light baryons.

\indent Finally, in Fig. 4 we have reported the results obtained for
the slopes $\rho_{(i)}^2 \equiv -[d\xi^{(i)}(\omega) / d\omega]_{\omega
= 1}$ ($i = 0, 1$) at the zero-recoil point as a function of the
parameter $\beta$ in the saturation region $\alpha > \alpha_{sat}$.
The slope $\rho_{(1)}^2$ results to be systematically larger than
$\rho_{(0)}^2$ and both of them are monotonically decreasing function
in a wide range of values of $\beta$. In the limit $\beta \to \infty$
the slope $\rho_{(1)}^2$ approaches the value $\simeq 2.1$, while
$\rho_{(0)}^2$ monotonically decreases up to $\simeq 1.1$, i.e. it
becomes very close to the value of the slope $\rho^2$ of the $IW$ form
factor for heavy-light mesons \cite{SIM96}. The values of
$\rho_{(0)}^2$, which can be found in literature, are spread in a
quite large range, namely: $\rho_{(0)}^2 = 3.7$ \cite{GK93}, $2.4$
\cite{SZ93}, $1.3$ \cite{JMW96}, $1.15$ \cite{YAK96} and
$1.2_{-1.1}^{+0.8}$ \cite{UKQCD}.

\indent In conclusion, the space-like elastic form factors of baryons
containing a heavy quark have been investigated within the light-front
constituent quark model of Refs. \cite{CAR,SIM96} in the limit of
infinite heavy-quark mass, adopting a gaussian-like ans\"atz for the
three-quark wave function. It has been found that the Isgur-Wise
functions, corresponding both to a spin-0 and a spin-1 light
spectator pair, depend strongly on the baryon structure, being sharply
different in case of diquark-like or collinear-type configurations in
the three-quark system. Moreover, it has been shown that relativistic
effects lead to a saturation property of the calculated Isgur-Wise
form factors as a function of the baryon size. Finally, our
predictions have been compared with those of different models as well
as with recent predictions from $QCD$ sum rules and lattice $QCD$
simulations; the latter ones seem to suggest the dominance of
collinear-type configurations, in which the heavy-quark is sitting
close to the center-of-mass of the light quark pair.

\newpage

\begin{center}

{\bf Figure Captions}

\end{center}

\vspace{2cm}

\indent {\bf Figure 1.} The form factor $\xi^{(0)}(\omega)$ (Eq.
(\ref{eq:csi0})), evaluated using the $LF$ wave function
(\ref{eq:LF_wf}) and adopting the gaussian ans\"atz (\ref{eq:gaussian})
for the radial function, as a function  of the recoil $\omega$ for
various values of the parameter $\alpha$ (Eq. (\ref{eq:alpha})).
The dot-dashed, dotted, dashed, long-dashed and solid lines correspond
to $\alpha = 0.1, 0.2, 0.3, 0.4$ and $1.0 ~ GeV$, respectively. The
light $CQ$ masses are $m_q = m_{q'} = 0.220 ~ GeV$ and the parameter
$\beta$ (Eq. (\ref{eq:beta})) is fixed at the value $\beta = 2$. 

\vspace{1cm}

\indent {\bf Figure 2.} Diquark-like (a) and collinear-type (b)
three-quark configurations, corresponding to the limiting cases $\beta
<< 1$ and $\beta >> 1$, respectively (see text). The full dot
represents the heavy-quark, while a spectator $CQ$ mass ratio $\eta =
m_q / m_{q'} = 1$ is understood.

\vspace{1cm}

\indent {\bf Figure 3.} The $IW$ form factors $\xi^{(0)}(\omega)$ (a),
$\xi^{(1)}(\omega)$ (b) and $\zeta^{(1)}(\omega)$ (c), given by Eqs.
(\ref{eq:csi0})-(\ref{eq:zeta1}) calculated in our $LF$ $CQ$ model
adopting the gaussian ans\"atz (\ref{eq:gaussian}), as a function of
the recoil $\omega$ for various values of the parameter $\beta$ (Eq.
(\ref{eq:beta})). The dot-dashed, dotted, dashed, long dashed and
solid lines correspond to $\beta = 0.5, 1.0, 2.0, 3.0$ and $10.0$,
respectively. The value of $\alpha$ (Eq. (\ref{eq:alpha})) is always
in the saturation region $\alpha > \alpha_{sat}$ (see Eq.
(\ref{eq:alpha_sat})), while the value of the spectator $CQ$ mass ratio
is $\eta = m_q / m_{q'} = 1$. In (a) the results of the diquark model
of Ref. \cite{GK93} (with $\varepsilon = 0.6 ~ GeV$ and $b = 1.77 ~
GeV^{-1}$), the $MIT$ bag model of Ref. \cite{SZ93}, the Skyrme model
of Ref. \cite{JMW96} and the $QCD$ sum rule of Ref. \cite{YAK96}
(corresponding to $\rho_{(0)}^2 = 1.15$) are reported by the full
triangles, dots, diamonds and squares, respectively. The open dots and
squares correspond to recent lattice $QCD$ simulations \cite{UKQCD} of
axial and vector form factors, respectively. In (b) the long dashed and
solid lines are indistinguishable. In (c) the error bars on the solid
line indicate the uncertainty related to the Monte-Carlo integration
procedure in the calculation of the form factor $\zeta^{(1)}(\omega)$
(in case of $\xi^{(0)}(\omega)$ and $\xi^{(1)}(\omega)$ the
uncertainty is smaller by one order of magnitude).

\vspace{1cm}

\indent {\bf Figure 4.} The slopes $\rho_{(i)}^2 \equiv
-[d\xi^{(i)}(\omega) / d\omega]_{\omega = 1}$ ($i = 0, 1$) at the
zero-recoil point, predicted by our $LF$ $CQ$ model using the
gaussian ans\"atz (\ref{eq:gaussian}), as a function of the parameter
$\beta$ (Eq. (\ref{eq:beta})). The solid and dashed line correspond to
$i = 0$ and $1$, respectively. The value of the parameter $\alpha$
(Eq. (\ref{eq:alpha})) is always in the saturation region $\alpha >
\alpha_{sat}$ (see Eq. (\ref{eq:alpha_sat})), while the value of the
spectator $CQ$ mass ratio is $\eta = m_q / m_{q'} = 1$.

\newpage

\begin{figure}[htb]

\vspace{-1cm}

\epsfxsize=16cm \epsfig{file=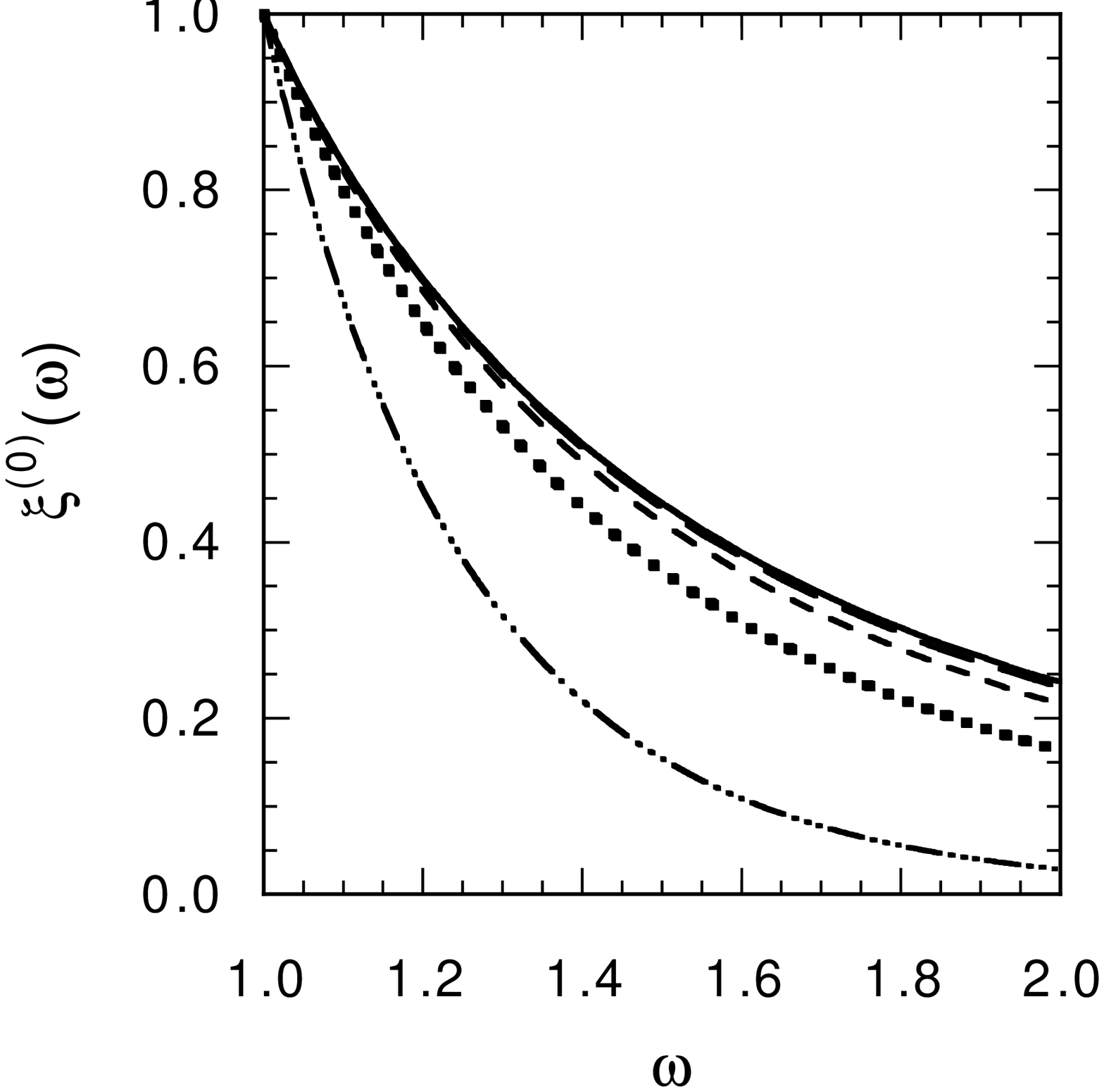}

\vspace{-5cm}

\centerline{\large{\bf Figure 1}}

\end{figure}

\newpage

\begin{figure}[htb]

\vspace{-1cm}

\epsfxsize=16cm \epsfig{file=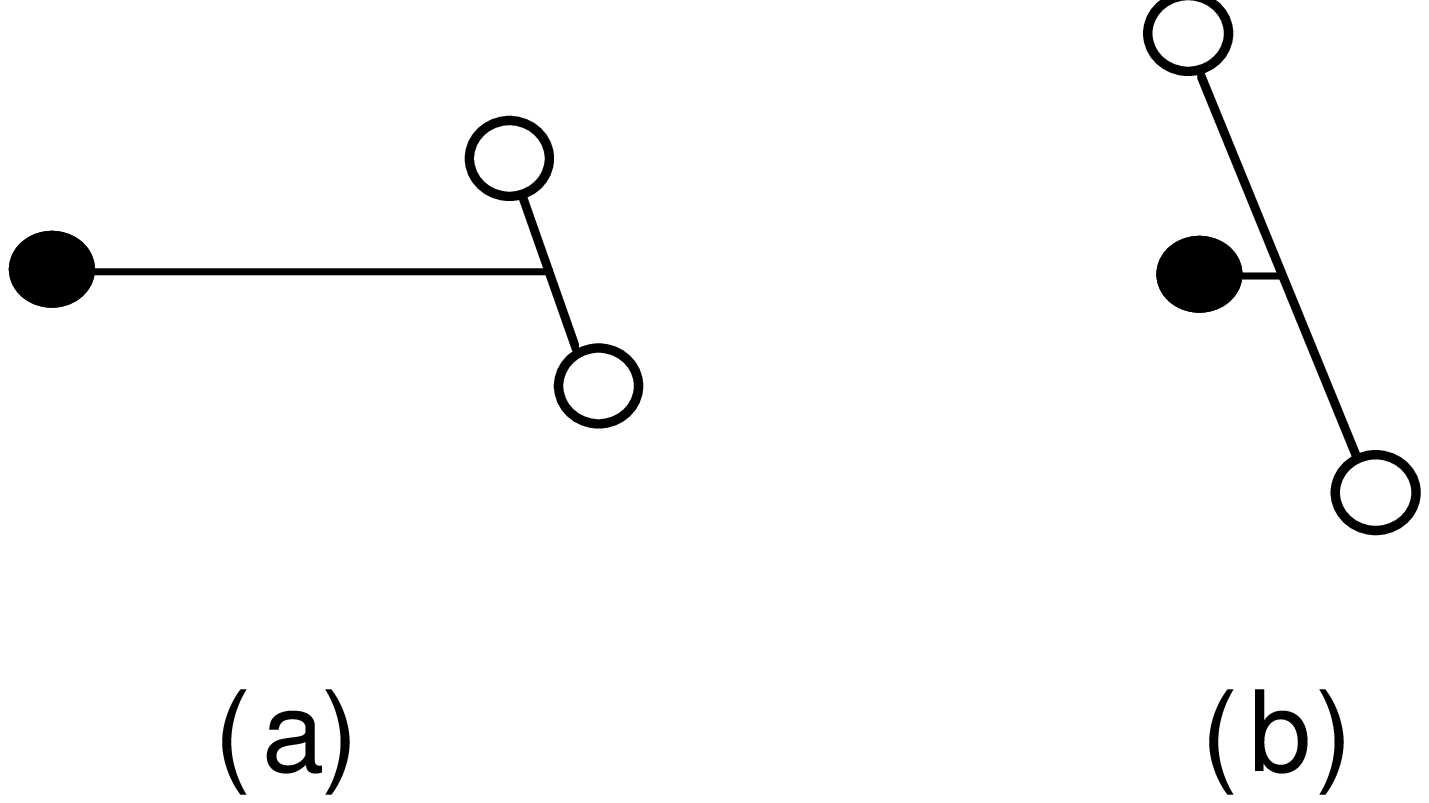}

\vspace{-10cm}

\centerline{\large{\bf Figure 2}}

\end{figure}

\newpage

\begin{figure}[htb]

\vspace{-1cm}

\epsfxsize=16cm \epsfig{file=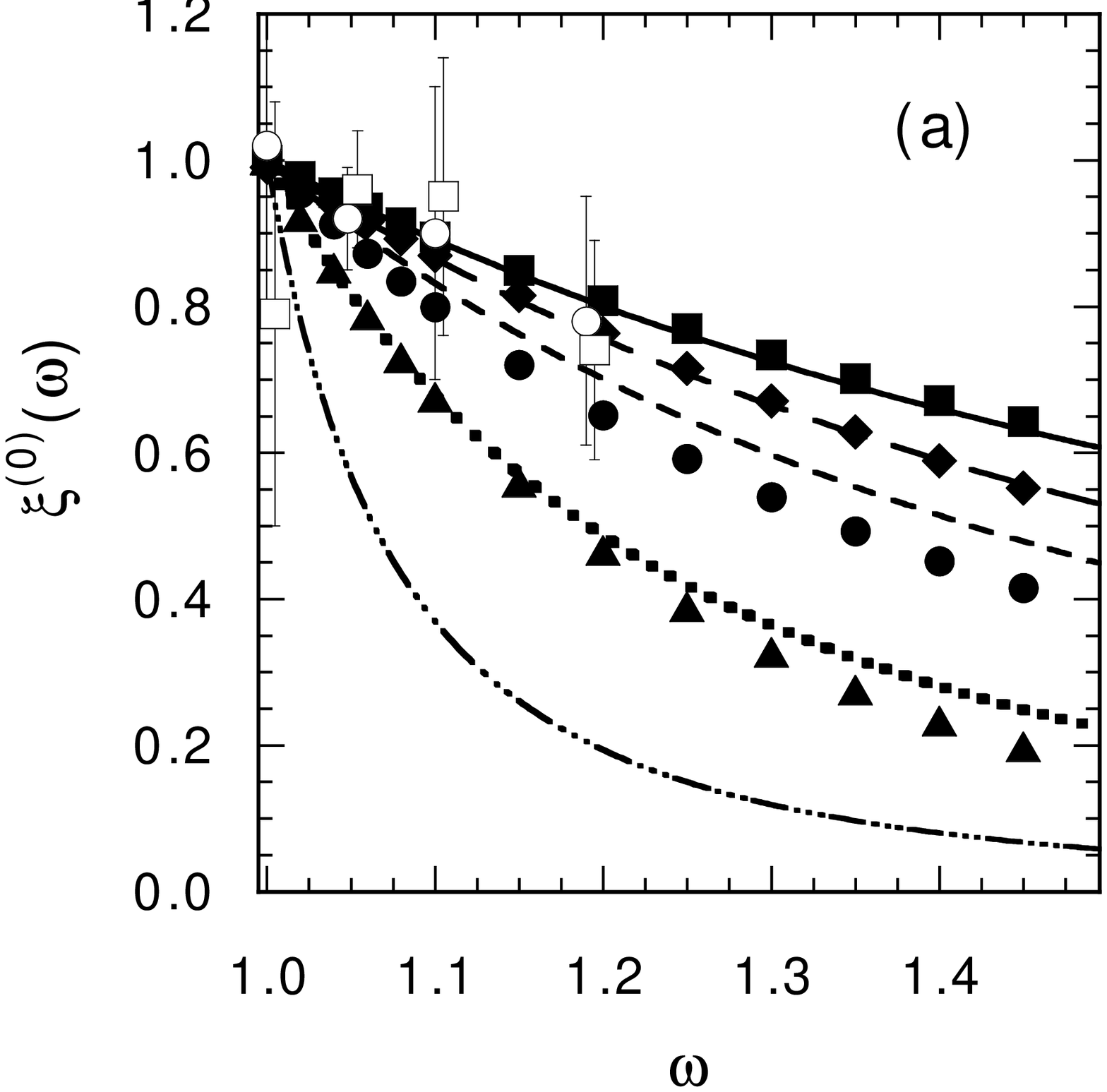}

\vspace{-5cm}

\centerline{\large{\bf Figure 3(a)}}

\end{figure}

\newpage

\begin{figure}[htb]

\vspace{-1cm}

\epsfxsize=16cm \epsfig{file=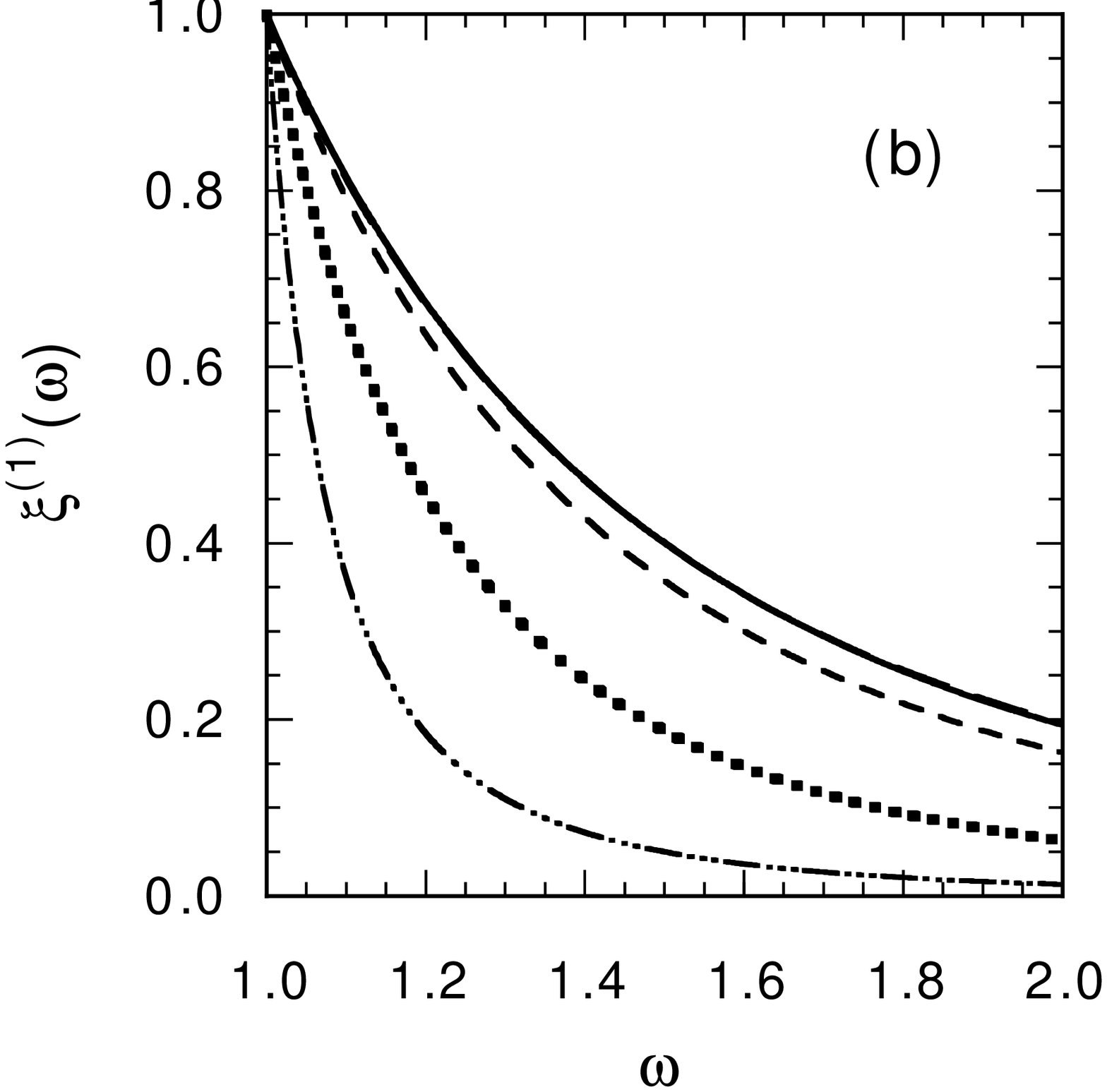}

\vspace{-5cm}

\centerline{\large{\bf Figure 3(b)}}

\end{figure}

\newpage

\begin{figure}[htb]

\vspace{-1cm}

\epsfxsize=16cm \epsfig{file=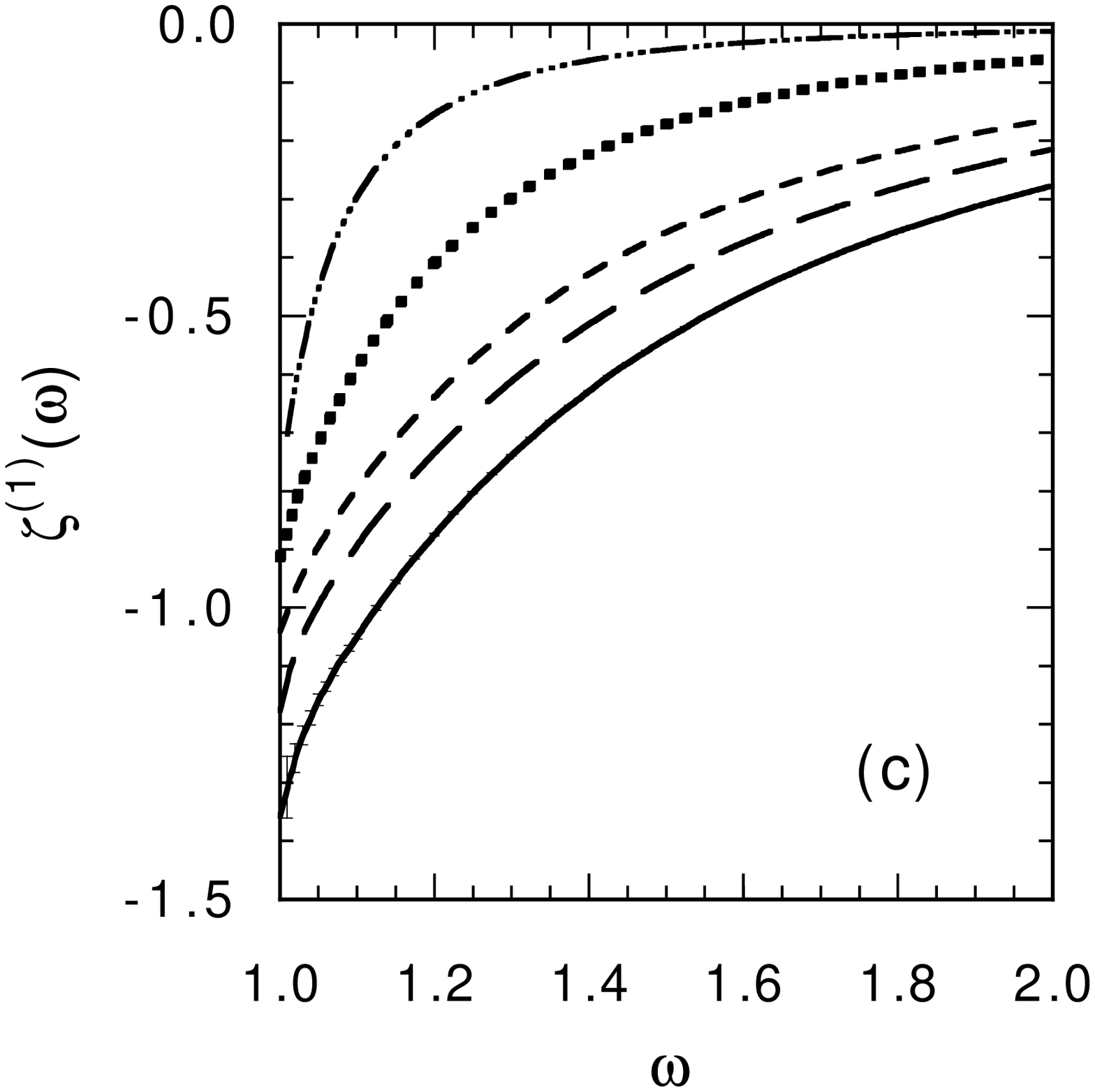}

\vspace{-5cm}

\centerline{\large{\bf Figure 3(c)}}

\end{figure}

\newpage

\begin{figure}[htb]

\vspace{-1cm}

\epsfxsize=16cm \epsfig{file=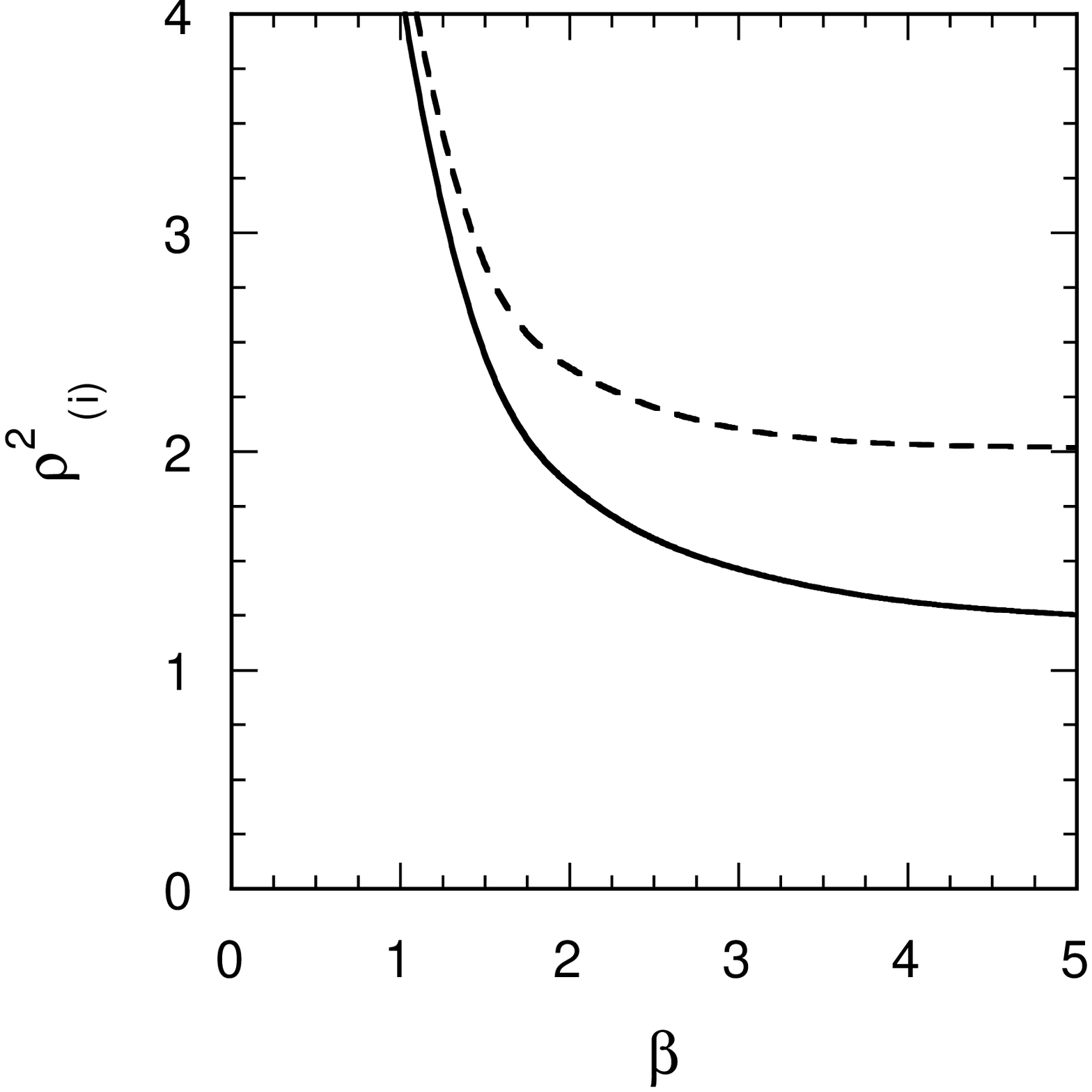}

\vspace{-5cm}

\centerline{\large{\bf Figure 4}}

\end{figure}


\begin{thebibliography}{99}

\bibitem{IW} (a) N. Isgur and M.B. Wise: Phys. Lett. {\bf B232} (1989)
 113; (b) {\em ib.} {\bf B237} (1990) 527; (c) Nucl. Phys. {\bf B348}
 (1991) 276.

\bibitem{HQET} D. Politzer and M.B. Wise: Phys. Lett. {\bf 208B}
 (1988) 504. M.B. Voloshin and M.A. Shifman: Sov. J. Nucl. Phys.
 {\bf 47} (1988) 511. H. Georgi: Phys. Lett. {\bf 240B} (1990) 447.

\bibitem{NEU94} For a review see, e.g., M. Neubert: Phys. Rep. {\bf
 245} (1994) 259, and references therein quoted.

\bibitem{CAR} F. Cardarelli et al: Phys. Lett. {\bf B332} (1994) 1;
 Phys. Lett. {\bf B349} (1995) 393; Phys. Lett. {\bf B357} (1995) 1;
 Phys. Lett. {\bf B359} (1995) 1; Few-Body Syst. Suppl. {\bf 8} (1995)
 345; Few-Body Syst. Suppl. {\bf 9} (1995) 267; Phys. Rev. {\bf D53}
 (1996) 6682; Phys. Lett. {\bf B371} (1996) 7; Phys. Lett. {\bf B397}
 (1997) 13; e-print archive nucl-th/9612063.

\bibitem{SIM96} S. Simula: Phys. Lett. {\bf B373} (1996) 193. I.L.
 Grach et al.: Phys. Lett. {\bf B385} (1996) 317; Nucl. Phys. {\bf
 B502} (1997) 227 (e-print archive hep-ph/9603369); Nucl. Phys. B
 (Proc. Suppl.) {\bf 55A} (1997) 84.

\bibitem{CW94} F.E. Close and A. Wambach: Nucl. Phys. {\bf B412}
 (1994) 169; Phys. Lett. {\bf B349} (1995) 207. See also A. Le Yaouanc
 et al.: Phys. Lett. {\bf B386} (1996) 304 and e-print archive
 hep-ph/9705324. 

\bibitem{VEGAS} G.P. Lepage: J. Comp. Phys. {\bf 27} (1978) 192.

\bibitem{CS} F. Cardarelli and S. Simula: to be published.

\bibitem{GK93} X.-H. Guo and P. Kroll: Z. Phys. {\bf C59} (1993) 567.
 See also X.-H. Guo and T. Muta: phys. Rev. {\bf D54} (1996) 4629.

\bibitem{SZ93} M. Sadzikowski and K. Zalewski: Z. Phys. {\bf C59}
 (1993) 677.

\bibitem{JMW96} E. Jenkins, A. Manohar and M.B. Wise: Nucl. Phys.
 {\bf B396} (1996) 38.

\bibitem{YAK96} A.G. Grozin and O.I. Yakovlev: Phys. Lett. {\bf
 B291} (1992) 441. O.I. Yakovlev: in Proc. of the $III$ German-Russian
 Workshop on {\em Heavy Quark Physics}, Dubna(Russia), May 20-22,
 1996, e-print archive hep-ph/9608348.

\bibitem{UKQCD} $UKQCD$ collaboration, K.C. Bowler et al.: e-print
 archive hep-lat/9709028.

\end{thebibliography}
\end{document}